# Is Einsteinian no-signalling violated in Bell tests?

Marian Kupczynski

Département de l'Informatique, Université du Québec en Outaouais (UQO),
Case postale 1250, succursale Hull, Gatineau, Quebec, Canada, J8X 3X 7

E-mail: marian.kupczynski@uqo.ca

## Abstract

Relativistic invariance is a physical law verified in several domains of physics. The impossibility of faster than light influences is not questioned by quantum theory. In quantum electrodynamics, in quantum field theory and in the standard model relativistic invariance is incorporated by construction. Quantum mechanics predicts strong long range correlations between outcomes of spin projection measurements performed in distant laboratories. In spite of these strong correlations marginal probability distributions should not depend on what was measured in the other laboratory what is called shortly: non-signalling. In several experiments, performed to test various Bell-type inequalities, some unexplained dependence of empirical marginal probability distributions on distant settings was observed . In this paper we demonstrate how a particular identification and selection procedure of paired distant outcomes is the most probable cause for this apparent violation of no-signalling principle. Thus this unexpected setting dependence does not prove the existence of superluminal influences and Einsteinian no-signalling principle has to be tested differently in dedicated experiments. We propose a detailed protocol telling how such experiments should be designed in order to be conclusive. We also explain how magical quantum correlations may be explained in a locally causal way.

## Introduction

The violation of Bell-type inequalities [1, 2] was reported in several excellent experiments [3-8] confirming the existence of long-distance correlations predicted by quantum mechanics (QM). Several magical explanations of these correlations are given: quantum instantaneous correlations come from outside space time, they result from retro-causation (causation from the future to the past), they are due to superdeterminism (experimentalists have no freedom to choose the experimental settings) etc. A recent critical review of these ideas and extensive bibliography may be found in [9-11].

Several authors [12-53] arrived, often independently, to similar conclusions and explained rationally why Bell inequalities might be violated. Strangely enough these explanations have been neglected by the majority of the quantum information community and remain unknown to the general public.

In this paper we examine reported anomalies [54-58] which might suggest that *Einsteinian no-signalling* was violated in Bell tests. We propose new dedicated experiments allowing testing *no-signalling* in an unambiguous way.

However to make our paper self-contained we have to explain the origin of Bell tests, why they are important and how one may explain rationally the violation of Bell-type inequalities.

The paper is organized as follows.

In section 2 we compare the description of a measurement process in classical and quantum mechanics and explain what we understand by *realism* and *contextuality*.

In section 3 we discuss shortly EPR paradox and Bell inequalities.

In section 4 we explain how the violation of Bell-type inequalities may be explained in a locally causal way. We discuss also *contextuality loophole*, *freedom of choice loophole*, *coincidence-time loophole* and *sample homogeneity loophole*.

In section 5 we compare the ideal EPRB experiment with its experimental realisations and we explain why setting dependence of empirical marginal probability distributions does not necessarily mean that *Einsteinian no-signalling principle* is violated.

In section 6 we present in detail new protocols which should be used to test *no-signalling* in twin-photon beam experiments.

In section 7 we show how setting dependence of marginal distributions may be explained using a contextual hidden variable model [9-11] without violating *no-signalling*.

In section 8 we propose a new test of *no -signalling* in entanglement swapping experiment [6].

In section 9 we present some existing experimental data confirming *Einsteinian no-signalling*.

Section 10 contains additional discussion of topics treated in preceding sections and some conclusions. It is difficult to imagine how a superfast scalable quantum computer might be constructed using EPR pairs.

## 1. Local realism versus quantum contextuality

In everyday life we describe objects by various properties such as: size, form, colour, weight etc. In general these properties depend on the context of a measurement. For example if we have a metal rod its length depends on variations of the ambient temperature, its weight depends on the place on earth we measure it. However if we fix experimental context: same temperature and the same place on earth its length and its weight do not depend on the order chosen to measure these compatible physical observables. Measurements of length and weight have a limited precision but in classical physics it is assumed that this precision may be always improved.

In classical mechanics (CM) an important idealisation is a material point. Due to large distances a motion of planets around the Sun can be modeled as a motion of material points. A relative motion of a material point with respect to an observer may be determined using successive (non-disturbing and accurate) measurements of its position $\mathbf{r}$ and time t. The speed of light in vacuum does not depend on a speed of its source nor on a speed of an observer. Thus ($\mathbf{r}$, t) may be found using the radar method which consists on sending a light signal and measuring time when the reflected signal returns to the observer.

A couple ($\mathbf{r}$, t) is called an event and all events form 4 dimensional space–time. If the radar method is used different observers moving with constant relative velocities assign to the same event different coordinates ($\mathbf{r'}$, t') which are related by the transformations of Poincare group. A motion of a material point is represented as a line in this 4 dimensional Einsteinian space-time. If coordinates of events are related by the transformations of Galilean group we have a Newtonian space-time.

The concept of the space -time is necessary to describe our experiments and observations in classical physics. It is also necessary to describe macroscopic set-ups and outcomes obtained in experiments probing the properties of atoms and the properties of elementary particles. The space-time loses its empirical basis at the atomic scale. Nevertheless conservation laws, deduced from the symmetries of the space- time, such as the conservation of total energy-momentum and the conservation of total angular momentum remain valid in quantum mechanics (QM) and in quantum field theory (QFT).

In CM the disturbance of a measuring instrument on a physical system may be neglected. Thus *by realism in classical physics we understand that measuring instruments read pre-existing values of observables characterizing jointly a given physical system in a particular experimental context*. Some observables such as a rest mass and an electric charge are believed to be context-independent attributes of a physical system. A similar notion is *counterfactual definiteness* (CDF) according to which *a physical system is completely described by the values of some set of physical observables which have definite values even if we do not measure them*.

In QM, by contrast to CM, there exist incompatible physical observables. QM is teaching us that measuring instruments play an active role in a measurement process and cannot be neglected. Namely in *QM measurement outcomes are created in interaction of identically prepared physical systems with a whole experimental set-up giving contextual and complementary information about a state of a studied system*.

As we learn from Bertrand's paradox there is an intimate relation between a probabilistic model and a random experiment it wants to describe. One may say that probabilities are contextual

"properties" of random experiments [37, 59, 60]. QM gives probabilistic predictions for a statistical scatter of measurement outcomes. These predictions change if experimental contexts change thus QM is a *contextual theory*. More detailed discussion of intimate relation of probabilistic models with experimental protocols in relation to Bell tests may be found for example in [9].

Bohr claimed, without proving it, that complementary information gathered in different (often incompatible) experimental set-ups gives a complete description of individual physical systems [61]. He also believed that quantum probabilities are irreducible and that more detailed space-time description of quantum phenomena is impossible. Einstein never agreed with this claim [62,63] what will be the topic of the next section.

## 2. EPRB paradox and Bell inequalities

In 1935 Einstein, Podolsky and Rosen (EPR) [64] discussed so called EPR-experiments in which two physical systems interacted in the past and separated, but some of their properties remained strongly correlated. QM seemed to predict that by making a measurement on one of these systems one might predict with certainty the value of an observable describing the second system. By choosing different observables to be measured on the first system values of incompatible observables describing the second system might be deduced. EPR concluded that QM did not provide a complete description of individual physical systems.

The EPR paper stimulated discussions about the physical reality and the interpretation of QM, which have been continuing till now. A review of these discussions is outside of the scope of this paper. Our point of view may be found in [10,11,40,41]. Here we list some of our conclusions:

- In order to deduce the exact value of an observable describing the second system from the value of an observable measured on the first system much more information concerning this particular pair of physical systems is needed and is usually unavailable.

- To deduce values of incompatible observables describing the second system one has to use incompatible physical set-ups thus the different measurements cannot be done on the same pair of physical systems (*complementarity*, *contextuality*).

- Correlations between outcomes of distant experiments are never perfect and they are only known when all experimental outcomes are analyzed.

- Any sub- quantum description of a measurement process must be contextual by what we understand that it has to include measuring instruments.

Einstein believed that the apparent indeterminism of QM is caused by a lack of knowledge about states of physical systems. It means that pure quantum ensembles of identically prepared physical systems are in fact mixed statistical ensembles. Different members of these mixed ensembles are characterized by different values of some uncontrollable hidden variables $\lambda$ and all measurement outcomes are predetermined by values of these variables [62,63].

This picture seems to be particularly suited for explaining outcomes of the so called EPR-Bohm thought experiment [65] (EPRB). In EPRB pairs of electrons or photons are prepared by a source and spin projections , in various directions, are measured in distant laboratories by Alice and Bob (using modern terminology).

QM seems to predict that the probability of observing a *spin-up* or a spin-*down* outcome by Alice and Bob in any direction is ½. At the same time the outcomes obtained for each pair are 'perfectly' anti-correlated. We agree with Bell that such correlations *cry for explanation* [1,2].

The simplest explanation is that a source produces a mixed ensemble of pairs who have strictly anti-correlated spins and that Alice's and Bob's instruments are passively registering pre-existing values of spin projections. This mixed statistical ensemble resembles an ensemble of pairs of Bertelsmann's socks of different colour and sizes in which one sock is sent by Charlie to Alice and another to Bob. Thus Alice and Bob always receive socks of the same or of the different colours and sizes depending on how pairs of socks are prepared by Charlie.

If a *spin –up* and a *spin-down* outcomes are coded by 1 or -1 respectively we may say that in a setting (x,y) , in which spin polarizations are measured, outcomes registered by Alice and Bob are values of random variables $A_x = \pm 1$ and $B_y = \pm 1$ . In local realistic hidden variable models (LRHVM) expectation values of these random variables are written as:

$$E(A_x B_y) = \sum_{\lambda \in \Lambda} P(\lambda) A_x(\lambda_1) B_y(\lambda_2) \qquad (1)$$

where $\Lambda$ is a set of all "hidden variables", $\lambda = (\lambda_1, \lambda_2)$ describe different pairs produced by a source and $P(\lambda)$ is a probability distribution of them. Outcomes of spin projection measurements for all the settings are determined locally by values of $\lambda$. One may say that all prepared pairs have well defined spin projections in all directions before the measurement which does not modify them. The probabilistic model describes photons as they were Bertelsmann's socks in spite of the fact that the spin projections on different directions may not be measured at the same time and by no means can be considered photon's attributes.

Bell demonstrated that expectation values calculated using (1) obey Bell-inequalities which, for some settings (spin measurement directions) , are violated by the values predicted by QM [1]. Several other inequalities CHSH [66], CH [67], Eberhard's [68] were deduced and extensively tested. These inequalities hold also for so called stochastic hidden variable models (SHVM) [2,67] in which pair of photons is described as it was a pair of fair dices. Namely SHVM describes EPRB as a family of independent random experiments labelled by $\lambda = (\lambda_1, \lambda_2)$:

$$P(A_x = a, B_y = b) = P(a, b \mid x, y) = \sum_{\lambda \in \Lambda} P(\lambda) P(a \mid x, \lambda_1) P(b \mid y, \lambda_2) \qquad (2)$$

Each pair of photons produced by a source is described by $\lambda = (\lambda_1, \lambda_2)$ In order to estimate probabilities in (2) Alice and Bob have to repeat measurements on <u>the same pair of photons</u> what is impossible. A detailed discussion of various probabilistic models and experimental protocols in connection to Bell-tests may be found in [9]. The factorisation of the probabilities in (2):

$$P(a, b \mid x, y, \lambda_2) = P(a \mid x, \lambda_1) P(b \mid y, \lambda_2) \qquad (3)$$

is called Bell's locality condition. This is why the violation of Bell-type inequalities was called *quantum nonlocality* and seemed to prove the existence of mysterious instantaneous influences coming from outside space-time [69].

In SPCE various inequalities are violated as predicted by QM. Moreover setting-dependence (relative angle dependence) of correlation functions $E(A_x, B_y) = E(A, B \mid x, y)$ seems to agree reasonably with quantum predictions [3,4].

## 3. Causally local explanation of quantum correlations

Experiments with entangled photon pairs are complicated and may suffer from different experimental loopholes [70,71] which are believed to be closed in [6-8].

We personally do not doubt that various Bell-type inequalities are in fact violated. We explain shortly why they may be violated and how long-range correlations observed in spin polarisation experiments may be explained using a locally causal probabilistic model.

LRHVM try to describe various incompatible random experiments using a joint probability distribution on a unique probability space what is very restrictive, They neglect an active role of measuring instruments and suffer from <u>theoretical</u> *contextuality loophole* which cannot be closed [47-49]. If supplementary parameters describing measuring instruments are <u>correctly</u> introduced Bell-type inequalities may not be proven and probabilistic models contain enough free parameters to fit data from any spin polarization correlation experiment (SPCE).

The incorrectness of using a unique probability space to describe SPCE was pointed out by several authors [9-53] but the term *contextuality loophole* was explained and used for the first time by Theo Nieuwenhuizen in [48, 49].

Simple probabilistic models incorporating contextual hidden variables may be defined as in [11]:

$$E(A, B \mid x, y) = \sum_{\lambda \in \Lambda_{xy}} A_x(\lambda_1, \lambda_x) B_y(\lambda_2, \lambda_y) P_x(\lambda_x) P_y(\lambda_y) P(\lambda_1, \lambda_2) \qquad (4)$$

where $\Lambda_{xy} = \Lambda_1 \times \Lambda_2 \times \Lambda_x \times \Lambda_y$, $A_x(\lambda_1, \lambda_x)$ and $B_y(\lambda_2, \lambda y)$ are equal $0, \pm 1$. We see that by contrast to (1) random experiments performed in different settings are described using different parameter spaces $\Lambda_{xy}$ in agreement with QM and with Kolmogorov theory of probability. In

Kolmogorov theory each random experiment is described by its own probability space and its own probability measure. Only in very limited situations a unique probability space and a corresponding joint probability distribution may be used to describe a family of random experiments [9,16, 20,25,29,37,47-49,53].

Contextual models (4) use setting-dependent probability distribution of hidden variables. It is believed incorrectly that due to Bayes theorem such setting-dependence (called *freedom of choice loophole*) restricts experimenters' freedom of choice. The incorrectness of this reasoning is explained in detail in [11] thus the model (4) is consistent with *experimenters' freedom of choice*. "Nonlocal " correlations may be explained without evoking *quantum magic*.

Additional arguments against *quantum nonlocality* come from the fact that measurement outcomes (e.g. clicks on detectors) are events having space-time coordinates.

The description of the EPRB thought experiment, provided by QM, neglects the fact that in real experiments outcomes produced in distant locations carry registration time tags. These time tags are necessary in order to compare time series of events and to identify coincident outcomes of measurements performed on the members of the same EPR pair.

The identification of coincident outcomes of measurements on pairs of photons is a difficult task because of the presence of dark counts and laser drifts. Moreover some photons are lost and the interaction of a photon with a photomultiplier requires a finite elapse of time.

In 1986 Pascazio demonstrated [72] that by assuming particular time-delays one may construct local hidden variable models violating Bell inequalities. This has been called in the literature *coincidence-time loophole*. The importance of time coordinates and time delays was discussed in detail by Hess and Philipp [18-20] and Larsson and Gill [73].

The incorporation of time delays in the description of SPCE experiments allows a locally causal explanation of quantum correlations observed in SPCE. De Raedt, Michielsen and collaborators [74-79], assuming a reasonable dependence of time-delays on settings, simulated with success various SPCE. Using these simulations one may also obtain quasi perfect agreement with predictions of QM.

Moreover De Raedt, Michielsen and collaborators [77, 80,81] have shown that their models can be solved analytically for some choices of parameters and that for some sets of these parameters they can show that their models gives EXACTLY the same result as QM.

They also show that probabilistic models able to describe these computer simulation experiments do not suffer from *contextuality loophole* because probability distributions of hidden variables depend explicitly on experimental settings [77,79-81]. This is also consistent with Kochen-Specker theorem [82] which tells us that only contextual hidden variables may be consistent with quantum predictions.

Since various Bell-type inequalities were violated and a reasonable agreement of angular dependence of correlations with quantum predictions was reported one might ask where is a problem. However if raw experimental data are studied, in detail, several problems are noticed.

In 2007 Adenier and Khrennikov [54] analyzed data of Wehs et al.[4] and found anomalies which could not be explained using the *fair sample assumption*. The most troubling is the dependence (on distant settings) of marginal single count frequencies extracted from empirical joint probability distributions. It seems to be in conflict with quantum predictions and with *Einsteinian no-signaling* principle.

These anomalies were confirmed by an independent analysis by De Raedt, Jin and Michielsen [55, 56] and shown to be the result of the data analysis procedure to define coincidences. They also conclude that it is highly unlikely that these data are compatible with quantum theoretical description of EPRB. Similar anomalies were discovered by Adenier and Khrennikov [57] and by Bednorz [58] in Hensen et al. data [6].

Most recently using the work of Lin et al. [83] and of Zhang et al. [84] Liang and Zhang (unpublished results presented at FQMT2017) re-analyzed the data of [85] and reported that the probability (p-value) of observing some data points under the assumption of no-signalling was smaller than $3.17 \times 10^{-55}$. The result is derived assuming that the measurement settings were randomly chosen but it turned out that this assumption was not respected in the experiment of [85].

Liang and Zhang assumed that trials are independent and identically distributed. However it was also not checked carefully enough in the experiment [85]. We demonstrated with Hans de Raedt [86] that significance tests become meaningless if data samples are inhomogeneous. We concluded that if sample homogeneity was not tested carefully enough data suffer from *sample homogeneity loophole* (SHL) and statistical inference in terms of p-values may not be trusted. Unfortunately SHL was not or could not be closed in several Bell tests [43,87]. The indiscriminate use of p-values in different domains of science was also strongly criticized by Leek and Peng [88].

Nevertheless the anomalies reported above strongly suggest that new dedicated experiments testing Einsteinian no-signalling are needed. Einsteinian no-signalling is a fundamental physical law valid in classical physics and incorporated by construction in relativistic quantum filed theory (QFT) therefore it should not be violated.

In the next section we explain why dependence (on distant settings) of empirical marginal probability distributions, extracted from empirical joint-probability distributions in SPCE, is not synonymous to the violation of *Einsteinian no-signalling*.

## 4.  Conventional tests of no-signalling in Bell tests

It is well known in statistics that the existence of correlations between the outcomes of distant experiments does not require hidden influences between experimental set-ups or communication between Alice and Bob.

One might ask why one should have doubts about the validity of *Einsteinian no-signalling*.

The speculations about mysterious quantum non-locality and violation of no-signalling were inspired by Bell. Bell did not understand the limitations of LRHVM and of SHVM and he thought that the violation of his inequalities could only be explained by the existence of *superluminal influences* between distant experimental set-ups or because of *superdeterminism* (which he rejected), More detailed discussion of this point is given in [11].

To avoid signalling Alice and Bob have to choose their settings randomly. Setting choices and measurements performed using a given pair of the settings have to be space-like events in order to close the so called *locality loophole*.

We do not believe that there are causal or superluminal influences between distant experimental set-ups thus there should be no significant difference in the outcomes whether the settings are chosen randomly or not, whether the *locality loophole* is closed or not. This conjecture may and should be tested using standard statistical methods in dedicated experiments.

Moreover since the directions defining the settings are not sharp mathematical vectors QM does not predict perfect correlations or anti-correlations for even maximally entangled quantum state [9,36,39] what was confirmed by experimental data.

In the EPRB thought experiment there are no losses of pairs, the outcomes for each pair are coded by the values of two random variables (A ,B) where A = ±1 and B = ±1 . The experimental run is described by unambiguous samples $S_A$ = { $a_1,…,a_n$ } , $S_B$ = { $b_1,…,b_n$ } and $S_{AB}$ = { $a_1 b_1,…, a_n b_n$}. Using these samples one may estimate P(A = a|x, y), P(B = b|x, y) and P(A = a, B = b|x, y). Here  P(A = a|x, y) and P(B = b|x, y) are standard marginal probability distributions extracted from the joint probability distribution  P(A = a, B = b|x, y).

In the EPRB thought experiment the way how pairs are emitted is irrelevant and the quantum state vector does not depend on space coordinates and on time. It represents a stationary flow of pairs of entangled spins which are measured by Alice and Bob.

Settings of Alice and Bob are chosen at random at the moment when pairs arrive to distant measuring stations. If there are no superluminal influences marginal probability distributions:

$$P(A = a \mid x, y) = \sum_b P(A = a, B = b \mid x, y) \qquad (5)$$

and

$$P(B = b \mid x, y) = \sum_a P(A = a, B = b \mid x, y) \qquad (6)$$

should not depend on distant settings. Thus for all a, b, x, y:

$$P(A = a \mid x, y) = P(A = a \mid x, y') = P(A = a \mid x) \qquad (7)$$

and

$$P(B = b \mid x, y) = P(B = b \mid x', y) = P(B = b \mid y). \qquad (8)$$

For a source preparing a singlet state : $P(A = a \mid x) = P(B = b \mid y) = 1/2$ for all x and y.

SPCE are not idealized EPRB experiments described above. In both pulsed and in continuously pumped twin-photon beam experiments pair emissions are governed by some stochastic process not described by QM. As we already mentioned above there are black counts, laser intensity drifts, photon registration time delays etc. Each detected click has its time tag. Even for a fixed pair of settings (x, y) time tags $t_a$ and $t_b$ of Alice's and Bob's clicks are different. One has to identify photons being members of the same emitted pair. Thus by contrast to the idealised EPRB experiments in real SPCE only the samples $S_A$ and $S_B$ are available and $S_{AB}$ needs to be constructed.

Correlated clicks are rare events and depend on a detailed protocol how time-dependent events registered on distant detectors are paired. A simplest but not unambiguous method is to call the outcomes coincident if $|t_a - t_b| \leq W/2$ where W is a width of some time-window. A detailed discussion how the data are gathered and the coincidences determined in different SPCE may for example found in [55,56,77, 79,89]. It may be annoying that different papers on Bell tests are using different notation.

Since different pairing procedures lead in general to samples $S_{AB}$ having different properties thus reported violations of equations (7) and (8) should be rather called *context dependence of marginal distributions* (CDMD) or the violation of *parameter independence* [70]. We will show in the next section that , contrary to the general belief, the violation of *parameter independence* does not prove the violation of *Einsteinian no-signalling*.

In order to compare QM predictions with experimental data of Weihs et al. the authors of [54-56] assumed that studied samples are simple random samples drawn from a statistical populations described by some quantum two- particle state which might be not maximally entangled. They excluded the existence of other sources sending some single polarized photons to Alice and Bob. Single photons coming from these photon sources might be also responsible for accidental coincidences and some reported anomalies in empirical probability distributions.

If one wants to obtain a reasonable agreement of experimental data with some quantum probabilistic model one has to use much more complicated density matrix and not a two particle pure state. For example Kofler, Ramelov, Giustina and Zeilinger [90] use the following density matrix in the standard V/H basis;

$$\rho_r = \frac{1}{1+r^2}\begin{bmatrix} 0 & 0 & 0 & 0 \\ 0 & 1 & Vr & 0 \\ 0 & Vr & r^2 & 0 \\ 0 & 0 & 0 & 0 \end{bmatrix} \qquad (9)$$

where 0 < r < 1 and a damping positive real factor V < 1. In reality V will be complex and there will be no elements of zero value in any realistic density matrix [90]. After corrections for dark counts, accidental coincidences, detector efficiencies a reasonable fit to the data was obtained for r = 0.297 and V = 0.965.

According to the authors the remaining deviations are due the fact that number of counts and correlations might be affected by even small imperfections in the alignment and calibration of polarizing beam splitter and dual wave-plates. The authors report also that the relative deviations of singles counts for the same setting (between –0.12 % and –0.36 %) are about two to five times larger than expected from purely statistical fluctuations. They do not consider these deviations as the indication of signalling but explain it correctly by laser intensity drifts which cause temporal variations of the pair production rate.

As we see it is difficult to extract a reliable information about correlations from photonic experiments. Thus we repeat again that the violation of the equations (7) and (8) may not be interpreted as the violation of Einsteinian *no-signalling*.

## 5. New protocols for testing no-signalling in twin-photon beam experiments

In simple words *no-signaling* means that Alice's outcomes are not significantly affected by what Bob is doing in his set-up and vice versa. It shoud not matter whether *locality loophole* is closed or not. In order to obtain meaningful results we have to create large samples in various experimental contexts and analyze these samples more in detail then it is done usually.

In twin-photon beam experiments one obtains two time-series of detector clicks. As we explained above to test no-signalling it is not necessary to define a pairing protocol of distant clicks. A pairing is only needed in order to estimate correlations and to perform the Bell test . Nevertheless we define in this section a particular data-pairing protocol in order to explain clearly why the setting dependence of empirical marginal distributions does not prove the violation of the no-signalling principle.

We suggest below an experimental protocol which should be followed by Alice and Bob.

1. With a source removed they should study time-series of dark counts on their detectors.

2. With a source in place but with PBS removed they should check whether their clicks are generated by the same or different Poisson processes or by some mixtures of Poisson processes. By gathering several sets of data they may detect laser drift or time dependent experimental noise. Having large samples they should study the homogeneity of their samples as it was recommended in [86].

3. With a source and PBS in place (on one or both sides) they should repeat point 2 for different fixed settings (x, y) and study time -series of observed <u>single counts</u>. Data of Alice should not show any significant differences no matter what Bob is doing and vice versa.

4. Next to study correlations between their outcomes they have to define a data -pairing protocol and estimate empirical joint probability distribution for a chosen setting (x, y). For example they may use fixed synchronized time -windows having varying width W. They should choose W in order to maximize number of events: a click on Alice's and /or Bob's detector or no click on both of them. The time-windows with multiple counts may be skipped. Please note that different data - pairing protocols are possible [4,55,56,89].

After proceeding as in point 4 outcomes of Alice and Bob are described now by a family of random variables (A' (W), B' (W)) where A' (W) = 0 or ±1 and B' (W) = 0 or ±1.

They may estimate P(A'(W) = a, B'(W) = b |x,y) and two corresponding marginal distributions P(A'(W) = a,|x,y) and P(B'(W) = b |x,y).

If *no- signalling* holds they are be unable to reject the hypotheses:

$$P(A'(W) = a \mid x, y) = P(A'(W) = a \mid x, y') = P(A'(W) = a \mid x) \qquad (10)$$

and

$$P(B'(W) = b \mid x, y) = P(B'(W) = b \mid x', y) = P(B'(W) = b \mid y). \qquad (11)$$

for a = 0 or ±1, b = 0 or ±1 and all x and y.

The equations (10) and (11) ( conditions for no- signalling) in general have not been tested carefully enough. Instead for each setting (x, y) samples containing couples of non-vanishing outcomes (a ≠ 0 and b ≠ 0) were selected and joint probability distributions P(A'(W) = a, B'(W) = b|x,y, a ≠ 0 and b ≠ 0) were estimated. From these empirical joint probability distribution the estimates of marginal probability distributions P(A'(W) = a|x,y, a ≠ 0 and b ≠ 0) and P( B'(W) = b|x,y, a ≠ 0 and b ≠ 0) were extracted.

These empirical marginal distributions were used to test the following equations:

$$P(A'(W) = a \mid x, y, a \neq 0, b \neq 0) = P(A'(W) = a \mid x, y', a \neq 0, b \neq 0,) \qquad (12)$$

and/or of

$$P(B'(W) = b \mid x, y, a \neq 0, b \neq 0) = P(B'(W) = b \mid x', y, a \neq 0, b \neq 0) \quad . \qquad (13)$$

Since for each setting we have different pairs of photons thus obtained post-selected samples of paired outcomes depend on time-tagged raw data obtained by Alice and Bob which are different in different settings. In conclusion these post-selected samples used to estimate these marginal distributions are setiing dependent and the violation of (11) and/or (12} means only CDMD and not the violation of *Einsteinian no-signalling*.

As we see the violation of (12) and (13) may be explained as the effect of setting-dependent post-selection of data. The degree of violation of (12) and (13) may depend on W. If there is no-signalling the equations (10) and (11) should hold for all reasonable values of W.

Let us note that

$$P(A'(W) = a \mid x, y, a \neq 0, b \neq 0) \neq P(A'(W) = a \mid x, a \neq 0) \qquad (14)$$

and

$$P(B'(W) = b \mid x, y, a \neq 0, b \neq 0) \neq P(B'(W) = b \mid y, b \neq 0) \qquad (15)$$

since $P(A'(W) = a \mid x, a \neq 0)$ and $P(B'(W) = b \mid y, b \neq 0)$ are estimated using all available data before the post-selection of events in which both $a \neq 0$ and $b \neq 0$. Of course if there is no-signalling these probabilities should not depend on what is done in a distant laboratory.

Therefore as we insisted above: testing of no-signalling consists on testing only the equations (10) and (11).

Let us now describe more in detail the experimental protocol to be used:

1. The raw data are clicks on Alice's and Bob's detectors. After each run of experiment with the settings (x,y) there are two samples of time-tagged data: $S_A(x,y) = \{(a_i, t_i) \mid i=1,\ldots n_x\}$ and $S_B(x,y) = \{(b_j, t'_j) \mid j=1,\ldots n_y\}$ where $a_i = \pm 1$ and $b_j = \pm 1$.

2. In principle one could test *no-signalling* directly by comparing samples $S_A(x, y)$ with $S_A(x, y')$ and $S_B(x, y)$ with $S_B(x', y)$. However it is easier to analyze these data using <u>fixed</u> synchronized windows of width W. For simplicity and in order to be consistent with a standard procedure we keep only time -windows in which there is no click at all or a click on one of Alice's or/and Bob's detectors. New obtained samples depend now on W: $S_A(x,y,W) = \{a_i \mid i=1,\ldots N_x\}$ and $S_B(x, y, W) = \{b_j \mid j = 1,\ldots N_y\}$ where $a_i = 0$ or $\pm 1$ and $b_j = 0$ or $\pm 1$.

3. The experiment should be repeated M times for each fixed setting (x,y) and using a fixed W. Alice has now M large samples $S_A(x,y,W)$ of the same size $N_x = N_1$. Similarly Bob has M large samples $S_B(x, y, W)$ with $N_y = N_2$. Keeping the same number of data points from different experimental runs minimizes the influence of laser intensity drifts. Having these M samples Alice and Bob should check the homogeneity of their samples as it was recommended in [43,86].

4. They should repeat the step 3 for different settings (x,y') and (x',y) obtaining each time M samples $S_A(x,y',W)$, $S_B(x,y',W)$, $S_A(x',y,W)$ and $S_B(x',y,W)$ and verify their consistency.

5. Finally if the hypothesis that the samples $S_A(x,y,W)$ and $S_A(x,y',W)$ are drawn from the same statistical population and that the samples $S_B(x,y,W)$ and $S_B(x',y,W)$ are drawn from the same statistical population for all x,x',y and y' may not be rejected then Alice and Bob may confidently conclude that there was *no signalling* between distant experimental set-ups.

The experimental protocol proposed above was never implemented. Instead in order to perform a Bell Test samples $S'_{AB}(x,y,W) = \{(a_i, b_i | i = 1,..,L_{xy}\}$, where $a_i = \pm 1$ and $b_i = \pm 1$, were post-selected. To estimate the marginal probability distributions for Alice's and Bob's two samples: $S'_A(x,y,W) = \{(a_i | i=1,...L_{xy}\}$ and $S'_B(x,y,W) = \{(b_i | i=1,...L_{xy}\}$ <u>were extracted from $S'_{AB}(x,y,W)$.</u>

The post-selection procedure, used to create these samples, is setting-dependent thus there is no a priori reason that the marginal probability distributions estimated using $S'_A(x,y,W)$ and $S'_A(x,y',W)$ should be the same. Also the marginal distributions estimated using $S'_B(x,y,W)$ and $S'_B(x',y,W)$ may be significantly different.

Let us note that if we extract from the samples $S_A(x,y,W)$ data items $a_i \neq 0$ we obtain a smaller sample $S_A(x,y,W|a \neq 0)$. No-*signalling* holds if $S_A(x,y,W|a \neq 0)$ and $S_A(x,y',W|a \neq 0)$ are consistent. In conclusion $S'_A(x,y,W)$ is not a fair sub-sample of $S_A(x,y,W|a \neq 0)$. Similarly $S'_B(x,y,W)$ is not a fair sub-sample of $S_B(x,y,W|b \neq 0)$.

In the experimental protocol similar to this used by Weihs et al. [4] one studies the influence of a global time shift $\Delta$ between synchronized time-windows in order to maximize the number of correlated clicks. Thus by using W and $\Delta$ we extract from the raw data samples: $S_A(x,y,W,\Delta)$ and $S_B(x,y,W,\Delta)$. <u>Each of these samples corresponds to a different random experiment labelled by W and $\Delta$.</u>

Each of these experiments may be described by its own Kolmogorov probability space. Similarly if we want to use QM to describe these experiments we cannot use one quantum state but different density matrices for each of them. A detailed analysis how outcomes of these different "post-selected random experiments" vary in function of W may be found in [55,56].

We see that QM does not provide a complete description of realistic SPCE: one may not use a unique quantum state to fit the data for different choices of W and $\Delta$.

It is however remarkable that using a particular W and Δ estimated angular dependence of correlation functions agrees reasonably well with correlations predicted by QM for the maximally entangled quantum state [3,4]

## 6. Setting dependence of marginal probabilities does not prove signalling

In section 3 we saw that if probabilistic contextual model (4) is used to describe SPCE then Bell-type inequalities may not be proven [11]. Let us show now that in this model *no-signalling* equations hold but the marginal probability distributions may be setting dependent.

Let us fix W and Δ and compare data obtained in different settings (x, y). These data may be described by some specific model (4) chosen ad hoc. According to this specific probabilistic model Alice's and Bob's expectations of random variables A and B are:

$$E(A \mid \text{x}) = \sum_{\lambda_1, \lambda_x, \lambda_2} A_x(\lambda_1, \lambda_x) P_x(\lambda_x) P(\lambda_1, \lambda_2) \qquad (16)$$

$$E(\text{B} \mid \text{y}) = \sum_{\lambda_1, \lambda_y, \lambda_2} B_y(\lambda_2, \lambda_y) P_y(\lambda_y) P(\lambda_1, \lambda_2). \qquad (17)$$

We see that E (A|x) does not depend on what Bob is doing and E (B|y) does not depend on what Alice is doing. By contrast to it marginal expectations extracted from non-vanishing paired data may be setting dependent:

$$E(A \mid \text{x}, \text{y}) = \sum_{\lambda \in \Lambda'_{xy}} A_x(\lambda_1, \lambda_x) P_x(\lambda_x) P_y(\lambda_y) P(\lambda_1, \lambda_2) \qquad (18)$$

$$E(\text{B} \mid \text{x}, \text{y}) = \sum_{\lambda \in \Lambda'_{xy}} B_y(\lambda_2, \lambda_y) P_x(\lambda_x) P_y(\lambda_y) P(\lambda_1, \lambda_2) \qquad (19)$$

where $\Lambda'_{xy} = \{\lambda \epsilon\ \Lambda_{xy} | A\ (\lambda_1, \lambda_x) \neq 0$ and B $(\lambda_2, \lambda_y) \neq 0\}$.

Similarly one obtains setting- independent <u>single count probability distributions</u>:

$$P(\text{a} \mid \text{x}) = \sum_{\lambda \in \Lambda_a} P_x(\lambda_x) P_y(\lambda_y) P(\lambda_1, \lambda_2) \qquad (20)$$

$$P(\text{b} \mid \text{y}) = \sum_{\lambda \in \Lambda_b} P_x(\lambda_x) P_y(\lambda_y) P(\lambda_1, \lambda_2) \qquad (21)$$

where $\Lambda_a = \{\lambda \epsilon\ \Lambda_{xy} | A(\lambda_1, \lambda_x) = \text{a }\}$ and $\Lambda_b = \{\lambda \epsilon\ \Lambda_{xy} | B(\lambda_2, \lambda_y) = \text{b}\}$.

We obtain also setting- dependent <u>marginal probability distributions</u>:

$$P(\text{a} \mid \text{x}, \text{y}) = \sum_{\lambda \in \Lambda'_a} P_x(\lambda_x) P_y(\lambda_y) P(\lambda_1, \lambda_2) \qquad (22)$$

$$P(\mathbf{b} \mid \mathbf{x}, \mathbf{y}) = \sum_{\lambda \in \Lambda'_b} P_x(\lambda_x) P_y(\lambda_y) P(\lambda_1, \lambda_2) \qquad (23)$$

where $\Lambda'_a = \{\lambda \epsilon \ \Lambda'_{xy} | A \ (\lambda_1, \lambda_x) = a\}$ and $\Lambda'_b = \{\lambda \epsilon \ \Lambda'_{xy} | B \ (\lambda_2, \lambda_y) = b\}$.

It is obvious that P (a|x) and P (b|y) obey *no-signalling* equations (7) and (8) but P (a|x, y) and P (b|x, y) may violate them.

Our model (4), computer simulations [74-81] and proposed experimental protocol to test *no-signalling* are specific to twin-photon beam experiments. The experimental protocol of Hensen et al. experiment [6] is completely different and is based on so called entanglement swapping.

## 7.  Tests of no-signalling in entanglement swapping experiments

Let us explain now the experiment [6] following closely [57] but keeping our notation.

In distant laboratories Alice and Bob own each a single nitrogen vacancy (NV) centre: electron spin in diamond. NV centres are prepared using laser excitation pulses and spins are rotated with microwaves. Rotations are switched fast enough such that they are space-like separated. Rotations define the measurement settings (x,y) .

The rotated NV centers for each setting (x, y) can be in a bright state (+1) in which many photons are emitted or in a dark state (-1) when no photon is emitted. Using synchronized clocks Alice's and Bob's photon detectors register a click (coded +1) or no click (-1). Each click has its own time-tag. The raw data obtained by Alice and Bob are two huge samples:
$S_A(x,y) = \{(a_i, t_i) \ | i = 1,…n_x\}$ and $S_B(x,y) = \{(b_j, t_j) \ | j = 1,…n_y\}$ where $a_i = \pm 1$ and $b_j = \pm 1$.

Photons emitted by the NV centres are not only sent to Alice's and Bob's detectors but they are also sent to overlap at a distant beam splitter located at C roughly midway between Alice and Bob. Each detected photon at C has also its time-tag. Coincidence detection at the output ports of this beam splitter indicates successful photon entanglement and the information of a successful entanglement between the NV centres. This event-ready signal is occurring within a precisely defined window after the arrival time of a sync pulse.

Nevertheless one may not avoid event-ready sampling since many event-ready detections are in fact due to reflections of the laser excitation pulses, instead of coming from photons emitted by the NV centres. These unwanted event-ready signals are filtered, not perfectly, by delaying the start time T of the coincidence window.

The separation of the spins by 1280 m defines a 4.27 µs time window during which the local events at A and B are space-like separated from each other and may be paired to estimate the correlations needed to test CHSH inequalities.

One has to play with the start time T and width of the coincidence window in order to eliminate photons coming from reflections and to keep enough photons coming from NV centres. The number of coincidences between outcomes of Alice and Bob measurements during the allowed time-window depends strongly on this offset time T. Other possible errors are: detector dark counts, microwave pulse errors and off-resonant excitation of the NV centers.

Since a probability of obtaining a valid event–ready signal is about $6.4 \times 10^{-9}$ thus the samples allowing testing CHSH inequalities are small and they depend on T. As we mentioned in the introduction Adenier and Khrennikov [57] and Bednorz [58] reported setting-dependence of marginal distributions what is usually considered to be the violation of *no-signalling* principle.

We explained why the violation of equations (12) and (13) in twin-photon beam experiments does not prove the violation of *Einsteinian no-signalling*. The experiment [6] is completely different and sampling does not depend on the settings (x,y). However to prove *no- signalling* it is sufficient to show that a statistic scatter of all Alice's outcomes does not depend significantly on what Bob is doing and vice versa.

Thus to test no-signalling we don't need to restrict our analysis to the data corresponding only to successful heralded entanglement swapping events. We may compare instead huge samples $S_A(x,y)$ with $S_A(x,y')$ and $S_B(x,y)$ with $S_B(x',y$ for different settings x, x',y , y '. Using synchronized fixed time windows of width W we may obtain binned samples :
$S_A(x,y,W) = \{a_i, |i = 1,...N_x\}$ ,$S_A(x,y') = \{a_j |j = 1,...N_x\}$, $S_B(x,y,W) = \{b_i, |i = 1,...N_y\}$ and $S_B(x,y',W) = \{b_j |j = 1,...N_y\}$ where $a = \pm 1$ and $b = \pm 1$.

If the hypothesis that the samples $S_A(x,y,W)$ and $S_A(x,y',W)$ are drawn from the same statistical population and that the samples $S_B (x,y,W)$ and $S_B (x',y,W)$ are drawn from the same statistical population for all x,x',y and y' may not be rejected, then Alice and Bob may confidently conclude that *no- signalling* principle was not violated in this experiment.

## 8. Some experimental data confirming no-signalling

Adenier and Khrennikov [54] analyzed raw data coming from two experiments of Weihs et al. [4]. The first experiment was a long distance experiment with fast switching (of the measurement directions), the second was a short distance experiment without fast switching. They analyzed two data files: *scanblue* and *bluesine* and found several anomalies.

A *scanblue* file contained data from the fast switching experiment in which a scan varying Alice's side modulator bias was performed, with both sides randomly fast switching between an equivalent +0 and +45 degrees angle. Modulating the bias from -100 to +100 was linearly equivalent to rotating the "corresponding PBS" from $-\pi/2$ to $+\pi/2$.

A *bluesine* file contained the data from a short distance experiment in which Alice varied the measurement setting angle while Bob kept the same setting.

Many things could have been different between the two experiments: different rate of pair production, better or worse alignment of the optical components, additional losses and depolarisation.

Nevertheless Alice's and Bob's single photon counts extracted from these different data files do not depend on the settings (angles between the optical axes of "corresponding polarisation beam splitters"). With permission of the authors we reproduce below, with small modifications, Fig. 3 and Fig.5b from [54] where these single counts are displayed.

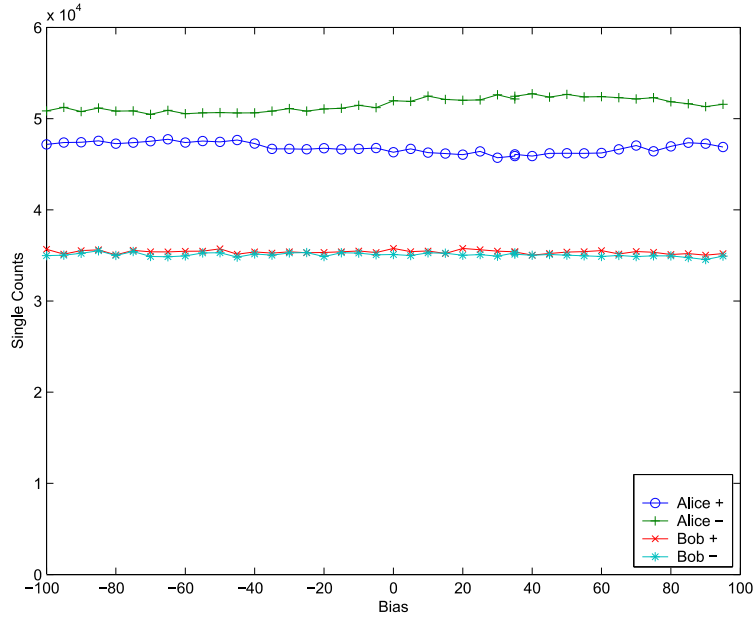

Figure 1: Single Counts (Switches 00, scanblue) [54]. Not original figure.

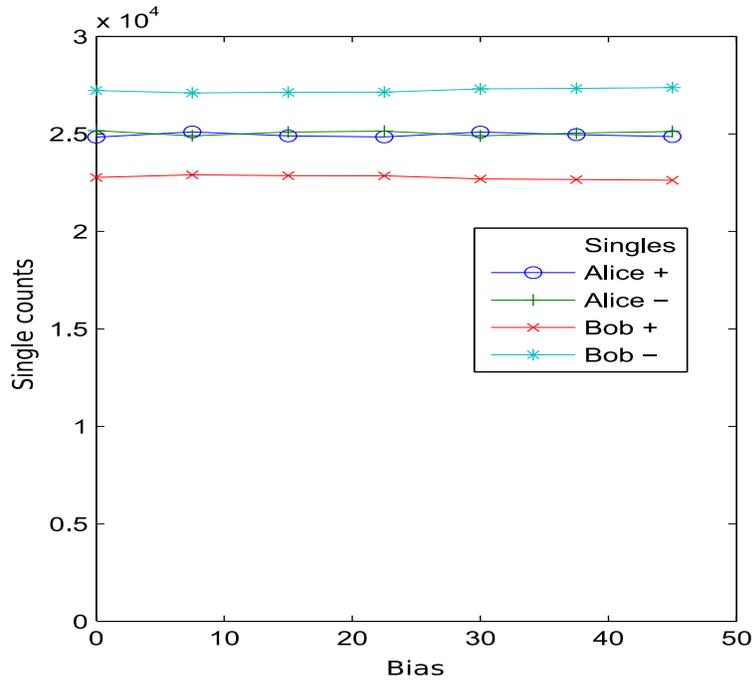

Figure 2: Single Counts (blusine) [54]. Not original figure.

According to our definition it seems to be a direct experimental confirmation of Einsteinian *no-signalling* principle in these experiments. Moreover it confirms our conjecture that *no*-signalling does not depend on closing the *locality loophole*.

By contrast to setting-independence of single counts authors demonstrated small but significant dependence of Alice's and Bob's empirical marginal probability distributions extracted from coincidence data [54]. This confirms our point of view, explained in detail in the sections 5-7, that setting- dependence of experimental marginal probability distributions may not be considered as the violation of *no-signalling* principle.

Let us notice that the data on the figure 1 seem to suggest that only Alice's detectors of 1 and -1 are unbalanced. The data on the figure 2 strangely enough seem to suggest that only Bob's detectors of 1 and -1 are unbalanced.

Single counts So$^A$ and So$^B$ for different settings ($\alpha i$, $\beta j$) are also available in the data from Giustina et al. experiment [7], Kofler et al .[90] compared these single counts in different setting:

|  | So$^A$ | So$^B$ |
|---|---|---|
| $\alpha 1, \beta 1$ | 1 526 617 | 1 699 881 |
| $\alpha 1, \beta 2$ | 1 522 865 | 4 515 782 |
| $\alpha 2, \beta 2$ | 4 729 369 | 4 507 497 |
| $\alpha 2, \beta 1$ | 4 735 046 | 1 693 718 |

Table 1: Single counts for different settings in Giustina et al. experiment [7].

As we may notice there is some setting dependence of Alice's and Bob's single counts So$^A$ and So$^B$. The differences between corresponding single counts are small but larger than statistical errors. They can easily be explained by the intensity drifts of a pump laser and do not prove signalling [90].

## 9. Discussion

In this paper we recall that long range correlations predicted by QM in experiments with entangled photons may be explained without violating locality and causality. Thus the violation of Bell-type inequalities does not allow to make any statements about local causality, non-locality of Nature, *superdeterminism*, experimenters' freedom of choice and completeness of QM.

By no means we want to say that one has to replace the description of these quantum experiments and phenomena by contextual probabilistic hidden variable model (4) or by the event-by-event computer simulation models [74-81]. These models are constructed to demonstrate that one does not need to evoke *quantum magic* to explain quantum correlations in SPCE.

Violations of Bell-type inequalities tell us only that we cannot explain incompatible experiments, in which the experimental instruments or environnement play an active role, using a standard joint probability distribution of some incompatible observables . This is why Bell inequlities are violated also in successive spin measurements as it was shown by Wigner [91]. They are also violated in several experiments in social sciences [30,31,92-94].

Bell inequalities may be even violated in classical mechanics. We present in [42] an ensemble of "entangled pairs "of metal balls colliding elastically on an air pillow. Hidden variables are speeds of balls and outcomes are strictly determined and coded as ± 1 by three different detectors A, B and C. In each repetition of the experiment, with balls having different initial speeds, one may use only one of these detectors on each of two sides of the experiment. It is impossible to use these 3 detectors at the same time on the same pair of balls, thus a joint probability distribution of associated random variables (A,B,C) does not exist. Bell inequalities may not be proven. By defining how the values of A, B and C are assigned to hidden variables we show that deduced correlations violate one of Bell inequalities.

It proves that there is nothing magical in entanglement and in "nonlocal" correlations predicted by QM. There are no hidden superluminal influences between members of entangled pairs or between distant experimental set-ups. The existence of such influences would contradict *Einsteinian no-signalling* principle which is incorporated in QM and in QFT.

Since some data showing the dependence of empirical marginal probabilities on distant setting have been reported we demonstrate in this paper that such dependence does not necessarily mean the violation of Einsteinian no- signalling principle.

In our opinion there is no reason for any (not necessarily superluminal) influences between distant experimental set-ups In this paper we propose new loophole free tests of *no-signalling* in SPCE. Our conjecture is that there will be no significant indication for the violation of no-signalling and it will not depend on whether *locality loophole* is closed or not whether the settings are chosen randomly or not.

EPRB experiment and Bell-type inequalities were proposed in order to test completeness of QM. Quantum description of EPRB does not contain time. By contrast to it in photon twin-beam experiments one has to use time-windows W and global time shift Δ to analyse the data and estimate the correlations. Post-selected experimental samples, used to perform Bell Tests, strongly depend on W, Δ and the settings. Standard quantum mechanical description of EPRB is unable to explain this dependence without additional assumptions. By contrast to it event-by-event locally causal simulations [74-79] are able to reproduce observed dependence of correlations on time-windows etc.

It seems to support Einstein's intuition that QM does not provide a complete description of individual physical systems. Similarly the failure of SHVM to explain quantum correlations seems to indicate that quantum probabilities are not irreducible and they may emerge from some more detailed description of quantum phenomena. There are several efforts to find such more detailed description let us cite here for example [95,96] where other references may be found.

Several years ago we pointed out that the completeness of QM should be tested by searching for some fine structures in time-series of data [34,35]. These fine structures are not predicted by QM and may be averaged out when data are analysed using a standard descriptive statistics. To find such structures one has to use sample homogeneity tests [86] and other tests specific to the study of experimental time-series [97-101].

It would be not easy to prove that QM is not predictably complete because we deal always in the experiment with mixed statistical populations depending on uncontrollable experimental factors. However even if one is not interested in completeness of QM sample homogeneity tests are essential in order to enable meaningful statistical inference from the data [86,87].

There is another worrying problem. Quantum probabilistic models are so flexible that using several parameters it is possible to fit various sets of data and never arrive to contradiction. In some sense QM may be unfalsifiable [102].

There are different exotic or less exotic interpretations of QM. We adopt, inspired by Einstein, Bohr and Ballentine [103] an unappealing contextual statistical interpretation [30,38,40,103-105]. In this interpretation there is no quantum magic. Wave functions are not attributes of individual physical systems which may be changed instantaneously [103]. It is meaningless to talk about a wave function of the Universe or to talk about retro-causation.

In this unappealing interpretation there are no mysterious influences and an outcome of Alice's measurement does not give immediate exact knowledge about Bob's outcome. One has to understand it well if one wants to construct a scalable superfast quantum computer using EPR pairs [106,107].

One has also to understand that Heisenberg inequalities and the impossibility of reproducing some quantum predictions by using joint probability distributions of incompatible variables does not mean, as Einstein said, that an electron and the MOON are not there when we do not look at them!

## Acknowledgements

The author would like to thank Andrei Khrennikov, Theo Nieuwenhuizen, Hans de Raedt, Kristel Michielsen, and Ethibar Dzhafarov for stimulating discussions. The author is also indebted to Guillaume Adenier for sending him graph files and for precious explanations.